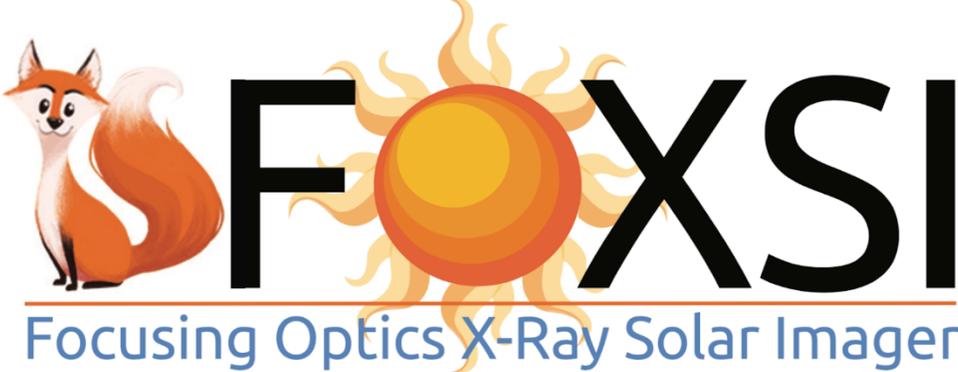

# Focusing Optics X-Ray Solar Imager

S. Christe (NASA Goddard Space Flight Center) et al. (see spreadsheet for complete list)

FOXSI is a direct-imaging, hard X-ray (HXR) telescope optimized for solar flare observations. It detects hot plasma and energetic electrons in and near energy release sites in the solar corona via bremsstrahlung emission, measuring both spatial structure and particle energy distributions. It provides two orders of magnitude faster imaging spectroscopy than previously available, probing physically relevant timescales (<1s) never before accessible to address fundamental questions of energy release and efficient particle acceleration that have importance far beyond their solar application (e.g., planetary magnetospheres, flaring stars, accretion disks). FOXSI measures not only the bright chromospheric X-ray emission where electrons lose most of their energy, but also simultaneous emission from electrons as they are accelerated in the corona and propagate along magnetic field lines. FOXSI detects emission from high in the tenuous corona, where previous instruments have been blinded by nearby bright features and will fully characterizes the accelerated electrons and hottest plasmas as they evolve in energy, space, and time to solve the mystery of how impulsive energy release leads to solar eruptions, the primary drivers of space weather at Earth, and how those eruptions are energized and evolve.

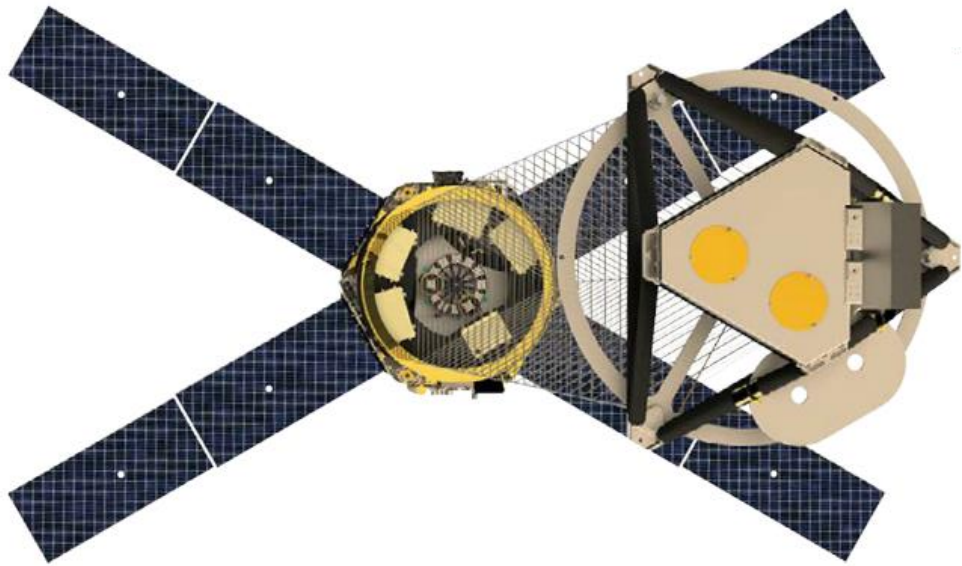

## 1.0 Introduction

How impulsive energy release leads to solar eruptions and how those eruptions are energized and evolve are vital unsolved problems in heliophysics. The standard model for solar eruptions (Figure 1) summarizes the current understanding of these events. Magnetic energy in the corona is released through drastic and rapid restructuring of the magnetic field via reconnection. Electrons and ions are then accelerated by poorly understood processes. Theories include contracting loops, merging magnetic islands, stochastic acceleration, and turbulence at shocks, among others[169,37,120,17,188]. Some accelerated particles can escape into the heliosphere to be observed as solar energetic particles (SEPs), and can have marked space-weather impacts. Other particles travel downward to the footpoints of magnetic loops in the dense chromosphere, where they deposit most of their energy. This energy rapidly heats chromospheric material to tens of millions of degrees, which then "evaporates" up into the corona. As magnetic fields reconnect, a coronal mass ejection (CME) can be released which can contain >$10^{12}$ kg of material traveling at speeds often exceeding 1000 km s$^{-1}$ and can also accelerate charged particles via shock-driven processes. These eruptions are the major force behind hazardous space weather, which can damage satellites, terrestrial power grids, and threaten human and robotic space explorers.

However, the fundamental physics underlying this well-established basic model is not understood. **Hard x-ray observations can probe all of the key regions of the standard model to provide breakthrough insights** (see black circles in Figure 1). These include two above-the-looptop sources, which bookend the reconnection region and are likely the site of particle acceleration and direct heating. The flux rope or CME core and hot plasma is also known to be a site of accelerated particles and hot plasma. Transport mechanisms affect the particles as they travel down the loop legs and are finally stopped by the dense chromosphere. This results in evaporation and hot flare loops. RHESSI has provided some of the most important observations of the standard flare model. However, the limited dynamic range and sensitivity afforded by the indirect imaging approach of RHESSI[78] and other instruments, such as the Hard X-ray Telescope (HXT) on Yohkoh[93] or the Spectrometer Telescope for Imaging x-rays (STIX) on Solar Orbiter[203], has not allowed key processes such as particle acceleration and direct heating to be studied together. This is because x-ray emission from the corona near the source of energy release and acceleration is relatively faint and cannot be separated by these instruments from the associated, but much brighter, chromospheric sources. In rare cases when RHESSI observed coronal sources, the foot point sources were either unusually weak or over the limb and hence not visible[97]. Focusing x-rays provides high dynamic range and sensitivity to finally investigate all key observational targets at the same time. This white paper discusses the science goals and objectives that can be achieved with a solar x-ray focusing optics instrument. It was originally proposed as a Small Explorer called the Focusing Optics X-ray Solar Imager (FOXSI) and was selected for a Concept Study Report. It was subsequently proposed as one of the instruments of a Medium Explorer concept called the Fundamentals of Impulsive Energy Release in the Corona Explorer (FIERCE). Both proposals received excellent ratings or both science and implementation.

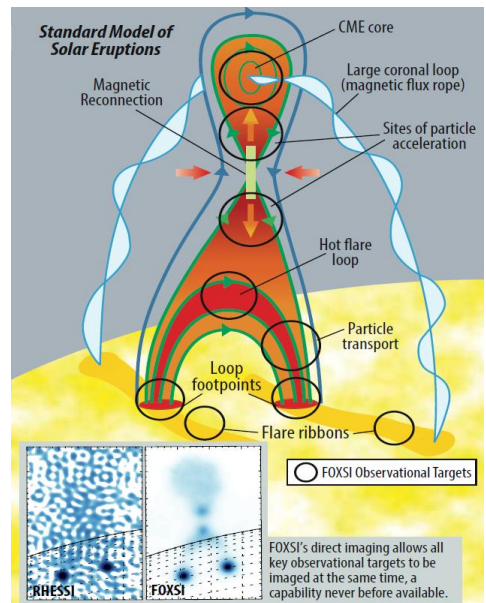

**Figure 1:** *The standard flare model. FOXSI measures spectra for every key region to determine the key physical mechanisms involved in solar eruptions.*

## 2.0 Technical Overview

The science described above requires observations of faint x-ray emission at sites of coronal energy release and particle acceleration as well as detecting the signatures of hot thermal plasmas. This can now be achieved by directly imaging hard x-rays in the 3 to 55 keV energy range with high spatial, spectral, and time resolution supported by fast soft x-ray observations with high spectral resolution. Direct imaging provides orders-of-magnitude improvements in dynamic range and sensitivity by focusing X-rays over large collecting areas onto small and fast pixelated detectors with low background. Combining high-angular resolution grazing-incidence hard X-ray focusing optics such as those routinely produced by the NASA Marshall Space Flight Center (MSFC) with fast pixelated solid-state detectors, such as the High energy X-ray imaging technology (HEXITEC) detectors provided by the Rutherford Appleton Laboratory (RAL)[165,157], provides direct imaging of the Sun with high resolution in energy, space, and time.

| Parameter | Performance |
|---|---|
| Number of optics | 2 |
| Energy Range | 3 to 55 keV |
| Total Angular Resolution | 8 arcsec (FWHM) |
| Imaging Dynamic Range | 1000:1 at >45 arcsec<br>20:1 at <20 arcsec |
| Effective Area | 55 cm$^2$ |
| Sensitivity | 0.2 photons cm$^{-2}$ |
| Imaging Time Resolution | 0.1 s |
| Energy Resolution | 0.8 keV |

**Table 1:** *FOXSI Instrument Performance*

Focusing x-rays requires long focal lengths. For the energy range discussed here, a deployable coilable boom (such as Northrop Grumman AstroMast already proven for this application on NuSTAR) easily accommodates a 14-m focal length. Unlike typical direct imagers at lower energies that integrate images over time, it is possible to record the energy, position, and time of arrival of each photon. This enables images to be produced on the ground without imposing excessively stringent requirements on pointing control or on the stiffness and alignment of the boom. This only requires precise knowledge of the instrument pointing at the time of arrival of each photon. This can be provided by a combination of a Fine Sun Sensor, to tell where the optics are pointed, and a boom metrology system, to tell where the detectors are with respect to the optics.

This technical approach has been matured through a series of successful flights on sounding rockets and balloons. Similar optics have flown on three FOXSI sounding rockets in 2012, 2014, 2018[101,103,104], after earlier development on five flights of the High Energy Replicated Optics balloon payload, which was also reconfigured and flown to observe the Sun as the HEROES mission in 2013. FOXSI also significantly benefits from experience learned from the Nuclear Spectroscopic Telescope Array (NuSTAR) mission, which pioneered the use of similar hard X-ray grazing-incidence optics on a satellite platform for astrophysics. A solar-optimized mission such as FOXSI can handle photon fluxes over 7 orders of magnitude, and has the angular resolution required to separate coronal and chromospheric sources in solar flares, unlike NuSTAR (see Table 1). Note that the quoted total angular resolution is a combination of the optics angular resolution (~7 arcsec), the pointing and metrology knowledge, and the discrete pixel size of 3.7 arcsec. Technology investments are suggested in a later section to improve this resolution.

## 3.0 Science

FOXSI addresses three science questions: 1) **How are particles accelerated at the Sun?** 2) **How do solar plasmas get heated to high temperatures?** And 3) **How does magnetic energy released on the Sun produce flares and eruptions?** These science questions address 7 out of 13 of the most important open issues identified in the last NASA Heliophysics Roadmap.

FOXSI measures the X-ray emission from electrons as they are accelerated in the corona and propagate along the magnetic field lines at the same time as the much brighter emission from electron that lose most of their energy in the chromospheric footpoints. Consequently, FOXSI can characterize the accelerated electrons and hottest plasmas as they evolve in energy, space, and time. FOXSI observes heated plasma and energetic electrons as close to the energy release site as possible, allowing key processes, such as particle acceleration and direct heating, to be accurately diagnosed. Furthermore, FOXSI measures x-ray spectra along the paths traveled by accelerated electrons from the corona to the chromosphere, obtaining previously-unavailable information about transport effects and associated heating.

### 3.1 How are particles accelerated at the Sun?

FOXSI provides systematic HXR observations from the particle acceleration site in the corona. Such observations, especially when combined with those of the much brighter chromospheric footpoint sources, will shed new light on how particles are accelerated at the Sun. FOXSI constrains acceleration models by observing x-ray spectra from accelerated electrons at locations and on time scales (<1 s) never before jointly accessible. FOXSI obtains imaging spectroscopy of thermal and non-thermal coronal and chromospheric x-ray sources simultaneously enabling the first systematic observations of non-thermal sources at and above the tops of flaring magnetic arcades even in the presence of much stronger x-ray footpoint sources. Some particle acceleration theories, such as those involving contracting and merging magnetic islands in the reconnection outflow[37,38], directly associate these sources with acceleration regions. Other theories predict acceleration sites in current sheets far above the flare[170] or in the looptops. The few RHESSI observations of double thermal sources above flare looptops suggest that the energy release site is between the two, possibly at termination shocks of reconnection outflows[171,118,119]. FOXSI observations of these high coronal regions enable direct testing of flare particle acceleration models. Additionally, FOXSI provides up to 2 orders of magnitude faster HXR imaging spectroscopy than previously available, probing physically relevant timescales (down to

~0.1 s) predicted by modeling, which show particle acceleration on timescales of 0.5 to 5 seconds[38,37].

The fraction of electrons accelerated out of the ambient Maxwellian velocity distribution is an essential constraint on acceleration models. The merging magnetic island theory[37] or acceleration by super-Dreicer electric fields in a reconnecting current sheet[117] can accelerate a large fraction of the available electrons, whereas models that invoke acceleration by large-scale sub-Dreicer electric fields[74] accelerate only a small fraction of the electrons. Interpretation of the spectra of these coronal sources has suggested that all electrons in the acceleration region are accelerated in some events[102,145] in a bulk energization process, despite many models suggesting that this is unphysical, e.g. due to the enormous return currents that would be produced. FOXSI will characterize coronal sources and the particle acceleration mechanism(s) involved (see Figure 2).

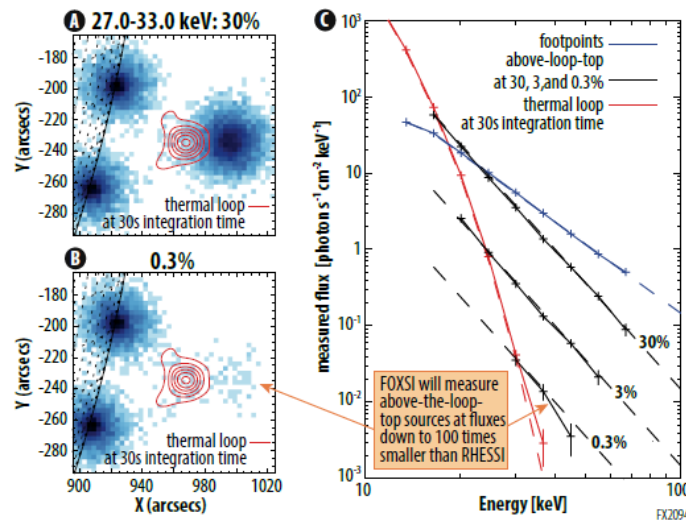

**Figure 2**: *FOXSI observes electrons as they are being accelerated in the above-the-loop-top acceleration region at the same time as the flare footpoints. Imaging simulations based on RHESSI observations of an M7.7 occulted flare (SOL2012-07-19T05:58) show the high dynamic range provided by FOXSI to probe potentially faint coronal sources.*

## 3.2 How do solar plasmas get heated to high temperatures

A recent major advance in solar physics is the development of reliable numerical codes for modeling the hydrodynamic response of the plasma to specified heating mechanisms, thereby predicting the resulting spectral line and continuum emission[24,2,16,155]. The application of these codes to observations provides understanding of the thermal evolution of the solar atmosphere during flares. Observations indicate that energy losses from accelerated electrons are a primary source of heating[48], which makes knowledge of the energy flux of these electrons critical. FOXSI provides simultaneous measurements of this energetic input and the evolution of the hot plasma. Accurate measurements of a flare's nonthermal energy content requires knowledge of the low-energy cutoff of the accelerated electron distribution. FOXSI's improved dynamic range will determine the nonthermal spectrum of footpoint sources down to lower energies even in the presence of a bright coronal thermal source. This will allow FOXSI to assess the total energy in accelerated electrons, which is a critical missing parameter in understanding the energetics of flares. Once electrons exit the coronal acceleration region, they suffer energy losses through Coulomb collisions, and possibly wave-particle interactions and return-current losses. The lost energy can substantially increase heating in the coronal part of the loop, at the expense of decreased energy input to the chromospheric footpoints. The electron distribution evolves with distance down the legs of the loop in a way that depends on the energy loss mechanisms ([Figure 3](#)). FOXSI will measure the x-ray spectrum at several locations along the flaring loop, enabling it to establish which energy loss mechanisms are dominant and where and how much energy is lost to the ambient corona.

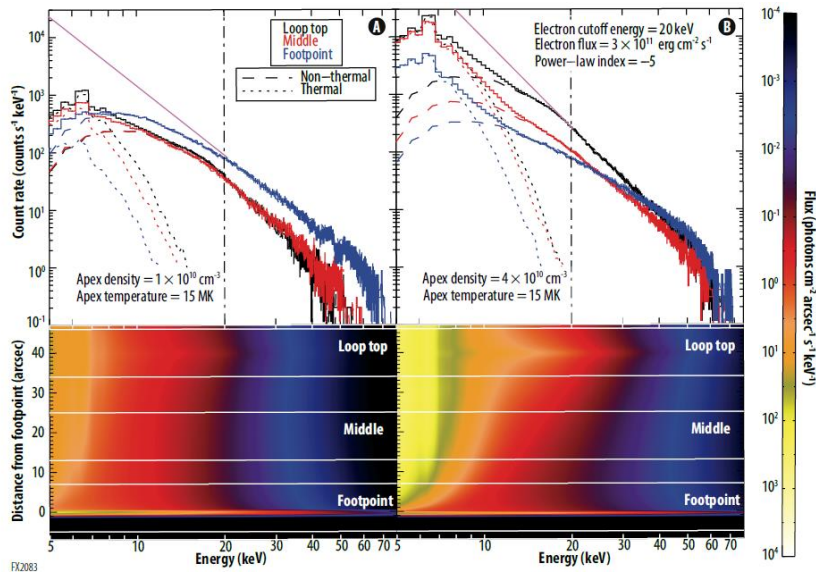

A growing body of evidence supports the idea that "direct" heating of flare plasma to high temperatures may be a signature of the reconnection process itself. FOXSI will provide a comprehensive, self-consistent test of coronal direct heating models as well as models for the co-evolution of accelerated electrons and thermal plasma in flares.

**Figure 3:** *FOXSI count-rate spectra calculated using RAYDN[204] for a GOES M5-class flare with two different apex loop densities. FOXSI observes significant differences between the two density cases in the spectra of each section of the loop. The different high-energy slopes reflect different emission mechanisms, thin-target emission from the loop and thick-target from the footpoints.*

With its high sensitivity, FOXSI will also place constraints on how flare-like processes such as nanoflares heat the corona. As a demonstration, the FOXSI-2 sounding rocket made the most direct measurement to date of high-temperature plasma in a solar active region[82]. This result provides strong support for impulsive heating above active regions. FOXSI will be able to distinguish between nanoflare heating models by observing the temperature distribution of active regions.

### 3.3 How does magnetic energy released on the Sun produce flares and eruptions?

With high sensitivity, dynamic range, and time cadence, FOXSI will observe the hot plasma inside the flux rope itself, which may erupt as a CME. Multiple studies[149,174,60] confirm the temporal correlation between the rate of reconnection, HXR emission, and CME acceleration in the low corona, which is known to be associated with geo-effectiveness[61].

CMEs can have hot cores of 10 MK[70], and direct HXR emission from non-thermal electrons has been observed from the cores of large CMEs[85,77,96,56]. Current HXR measurements of this phenomenon exist only for large events with occulted footpoints. FOXSI will observe HXR emission from CME cores and initiation in solar eruptive events at the same time as the bright flare coronal and footpoint sources (([Figure 4](#)). This will establish whether electron acceleration occurs in the current sheet below the CME, within the CME core itself, or elsewhere. The standard model of solar eruptive events shown in Figure E-1 is known to be missing more complex components of reconnection which can lead to additional accelerated electrons. One example is interchange reconnection, a process where emerging flux reconnects with open field lines. This leads to jets of hot plasma and accelerated electrons that can potentially escape the Sun. FOXSI's fast imaging spectroscopy with high dynamic range distinguishes between competing theories of the generation of jets and solar eruptive events.

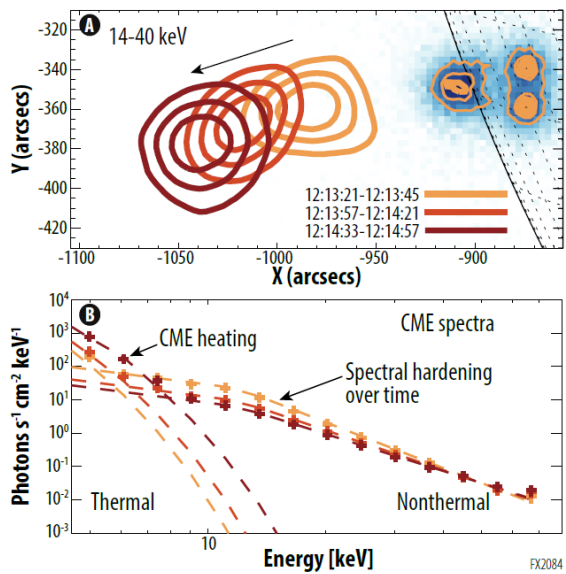

**Figure 4**: *FOXSI will observe CMEs as they are accelerated during a solar eruptive event. Based on a 2010 Nov. 3 CME event[56] with occulted footpoints observed by RHESSI, this simulation shows FOXSI observing heating and accelerated electrons inside a CME core at the same time as flare-accelerated electrons and heating at lower altitudes.*

## 5.0 Cost and Schedule

The cost and schedule for FOXSI both as an instrument and as a full Small Explorer mission have been extensively studied, modeled, and reviewed as part of the FOXSI Concept Study report and the FIERCE MIDEX proposal. FOXSI SMEX received a Low/Medium risk rating and was Category I. The individual instrument subsystems have high heritage (TRL≥6) and a flight instrument can be delivered in ~21 months for observatory integration including schedule reserve.

## 6.0 Technical Maturation Needs

The science goals and objectives described here can be achieved using existing capabilities. There are areas of investments that would significantly improve the science return of a mission including a FOXSI instrument. These include (1) high angular resolution x-ray optics and (2) fast photon-counting detectors with small pixels. Improved angular resolution for grazing incidence x-ray optics. Electro-formed nickel replicated optics provide a good balance of cost and angular resolution such that significant effective area is available through nesting shells. Higher angular resolutions (sub-arcsec) would enable resolving individual loops and footpoints and constrain the electron flux. This must be combined with high-flux small pixel solid-state x-ray detectors. Higher optical angular resolution must be combined with improved spatial resolution detectors at a fixed focal length. Detecting each individual photon, their energy and spatial location, at the high fluxes produced by large solar flares is not supported by many existing x-ray detectors. For the FOXSI design, our detector count rate requirement was at least 30,000 photons/s but this was combined with significant attenuation for large flares making it difficult to observe any emission below 10 keV.

## 6.0 Conclusion

FOXSI's primary science objectives are to understand the mystery of how impulsive energy release leads to solar eruptions, the primary drivers of space weather at Earth, and how those eruptions are energized and evolve. FOXSI solves this mystery by addressing fundamental questions of energy release and efficient particle acceleration that have importance far beyond their solar application (e.g., planetary magnetospheres, flaring stars, accretion disks). To obtain a comprehensive understanding of impulsive energy release, particle acceleration and transport, and heating in solar flares, FOXSI addresses three science questions: (1) How are particles accelerated at the Sun? (2) How do solar plasmas get heated to high temperatures? (3) How does magnetic energy released on the Sun produce flares and eruptions? This white paper provides only a summary of the science case made in the FOXSI SMEX proposal. Significant other scientific breakthroughs are possible when combining HXR observations by FOXSI with other instruments. A separate white paper on the Fundamentals of Impulsive Energy Release in the Corona Explorer (FIERCE) mission (by A. Shih et. al) has been submitted as well as one on the general topic of the need for focused, hard X-ray investigations of the Sun (by L. Glesener et. al). An instrument concept like FOXSI was recommended by the previous Heliophysics Decadal Survey as part of the SEE 2020 mission. The report further suggested that FOXSI "would be appropriate as an Explorer Mission and would provide tremendous new science."

To maximize the science return, a launch of a solar eruptive mission such as FOXSI SMEX or FIERCE must be timed with the maximum of the solar cycle. Given the 11-year solar cycle combined with the time required to formulate a mission concept, propose and implement, only one (or maybe two) such opportunity occurs every decade. For example, we began formulating our FOXSI SMEX mission concept in late 2015 and would have flown in Oct 2022, a 7 year delay. If an opportunity is missed, like it has been for Solar Cycle 25, the result is severe damage to the community of early career solar physicists which have been trained by the previous high energy solar mission, in this case, RHESSI. Significant attrition and recruitment difficulties should be expected if the community is asked to wait 11 years for the next opportunity. Supporting a vibrant and diverse high energy solar community is key to enabling scientific breakthroughs. Flight opportunities must be provided, and potentially guaranteed, to train and retain solar scientists whose work is tied to the solar cycle.

## 7.0 References

2      J. C. Allred, A. F. Kowalski, and M. Carlsson, "A Unified Computational Model for Solar and Stellar Flares," ApJ, vol. 809, no. 1, p. 104, Aug. 2015, https://doi.org/10.1088/0004-637X/809/1/104

16     S. J. Bradshaw and P. J. Cargill, "The Influence of Numerical Resolution on Coronal Density in Hydrodynamic Models of Impulsive Heating," ApJ, vol. 770, no. 1, p. 12, Jun. 2013, https://doi.org/10.1088/0004-637X/770/1/12

17     S. Brannon and D. W. Longcope, "Modeling Properties of Chromospheric Evaporation Driven by Thermal Conduction Fronts from Reconnection Shocks," ApJ, vol. 792, no. 1, p. 50, Sep. 2014, https://doi.org/10.1088/0004-637X/792/1/50

24     M. Carlsson and R. F. Stein, "Formation of Solar Calcium H and K Bright Grains," ApJ, vol. 481, no. 1, pp. 500–514, May 1997, https://doi.org/10.1086/304043

37     J. F. Drake, M. Swisdak, H. Che, and M. A. Shay, "Electron acceleration from contracting magnetic islands during reconnection," Nature, vol. 443, no. 7111, pp. 553–556, Oct. 2006, https://doi.org/10.1038/nature05116

38     J. F. Drake, M. Swisdak, and R. Fermo, "The Power-law Spectra of Energetic Particles during Multi-island Magnetic Reconnection," ApJ, vol. 763, no. 1, p. L5, Jan. 2013, https://doi.org/10.1088/2041-8205/763/1/L5

48     L. Fletcher, B. R. Dennis, H. S. Hudson, S. Krucker, K. J. H. Phillips, A. Veronig, M. Battaglia, L. Bone, A. Caspi, Q. Chen, P. T. Gallagher, P. T. Grigis, H. Ji, W. Liu, R. O. Milligan, and M. Temmer, "An Observational Overview of Solar Flares," Space Science Reviews, vol. 159, no. 1, pp. 19–106, Sep. 2011, https://doi.org/10.1007/s11214-010-9701-8

56     L. E. Glesener, S. Krucker, H. M. Bain, and R. P. Lin, "Observation of Heating by Flare-accelerated Electrons in a Solar Coronal Mass Ejection," ApJ, vol. 779, no. 2, p. L29, Dec. 2013, https://doi.org/10.1088/2041-8205/779/2/L29

60      N. Gopalswamy, H. Xie, S. Yashiro, S. Akiyama, P. Mäkelä, and I. G. Usoskin, "Properties of Ground Level Enhancement Events and the Associated Solar Eruptions During Solar Cycle 23," Space Science Reviews, vol. 171, no. 1, pp. 23–60, Oct. 2012, https://doi.org/10.1007/s11214-012-9890-4

61      N. Gopalswamy, P. Mäkelä, S. Akiyama, S. Yashiro, H. Xie, N. Thakur, and S. W. Kahler, "Large Solar Energetic Particle Events Associated with Filament Eruptions Outside of Active Regions," ApJ, vol. 806, no. 1, p. 8, Jun. 2015, https://doi.org/10.1088/0004-637X/806/1/8

70      I. G. Hannah and E. P. Kontar, "Multi-thermal dynamics and energetics of a coronal mass ejection in the low solar atmosphere," Astronomy and Astrophysics, vol. 553, p. A10, May 2013, https://doi.org/10.1051/0004-6361/201219727

74      G. D. Holman and S. G. Benka, "A hybrid thermal/nonthermal model for the energetic emissions from solar flares," ApJ, vol. 400, pp. L79–L82, Dec. 1992, https://doi.org/10.1086/186654

77      H. S. Hudson, "Observing coronal mass ejections without coronagraphs," Journal of Geophysical Research, vol. 106, no. 11, pp. 25199–25214, 2001, https://doi.org/10.1029/2000JA904026

78      G. J. Hurford, E. J. Schmahl, R. A. Schwartz, A. J. Conway, M. J. Aschwanden, A. Csillaghy, B. R. Dennis, C. Johns-Krull, S. Krucker, R. P. Lin, J. M. McTiernan, T. R. Metcalf, J. Sato, and D. M. Smith, "The RHESSI Imaging Concept," Solar Physics, vol. 210, no. 1, pp. 61–86, Nov. 2002, https://doi.org/10.1023/A:1022436213688

82      S.-N. Ishikawa, L. E. Glesener, S. Krucker, S. D. Christe, J. C. Buitrago-Casas, N. Narukage, and J. Vievering, "Detection of nanoflare-heated plasma in the solar corona by the FOXSI-2 sounding rocket," Nature Astronomy, pp. 1–4, Oct. 2017, https://doi.org/10.1038/s41550-017-0269-z

85      S. R. Kane, J. M. McTiernan, J. Loran, E. E. Fenimore, R. W. Klebesadel, and J. G. Laros, "Stereoscopic observations of a solar flare hard X-ray source in the high corona," ApJ, vol. 390, pp. 687–702, May 1992, https://doi.org/10.1086/171320

93      T. Kosugi, K. Makishima, T. Murakami, T. Sakao, T. Dotani, M. Inda, K. Kai, S. Masuda, H. Nakajima, Y. Ogawara, M. Sawa, and K. Shibasaki, "The Hard X-ray Telescope (HXT) for the SOLAR-A Mission," Solar Physics (ISSN 0038-0938), vol. 136, no. 1, pp. 17–36, Nov. 1991, https://doi.org/10.1007/BF00151693

96      S. Krucker, S. M. White, and R. P. Lin, "Solar Flare Hard X-Ray Emission from the High Corona," ApJ, vol. 669, no. 1, pp. L49–L52, Nov. 2007, https://doi.org/10.1086/523759

97      S. Krucker and R. P. Lin, "Hard X-Ray Emissions from Partially Occulted Solar Flares," ApJ, vol. 673, no. 2, pp. 1181–1187, Feb. 2008, https://doi.org/10.1086/524010

101     S. Krucker, S. D. Christe, L. E. Glesener, S. Mcbride, P. Turin, D. Glaser, P. Saint-Hilaire, G. Delory, R. P. Lin, M. V. Gubarev, B. D. Ramsey, Y. Terada, S.-N. Ishikawa, M. Kokubun, S. Saito, T. Takahashi, S. Watanabe, K. Nakazawa, H. Tajima, S. Masuda, T. Minoshima, and M. Shomojo, "The Focusing Optics X-ray Solar Imager (FOXSI)," SPIE Optical Engineering + Applications, 2009, vol. 7437, pp. 743705–743705–10, https://doi.org/10.1117/12.827950

102     S. Krucker, H. S. Hudson, L. E. Glesener, S. M. White, S. Masuda, J. P. Wuelser, and R. P. Lin, "Measurements of the Coronal Acceleration Region of a Solar Flare," ApJ, vol. 714, no. 2, pp. 1108–1119, May 2010, https://doi.org/10.1088/0004-637X/714/2/1108

103     S. Krucker, S. D. Christe, L. E. Glesener, S.-N. Ishikawa, S. McBride, D. Glaser, P. Turin, R. P. Lin, M. V. Gubarev, B. D. Ramsey, S. Saito, Y. Tanaka, T. Takahashi, S. Watanabe, T. Tanaka, H. Tajima, and S. Masuda, "The Focusing Optics X-ray Solar Imager (FOXSI)," SPIE, 2011, vol. 8147, pp. 814705–814705–14, https://doi.org/10.1117/12.895271

104     S. Krucker, S. D. Christe, L. E. Glesener, S.-N. Ishikawa, B. D. Ramsey, M. V. Gubarev, S. Saito, T. Takahashi, S. Watanabe, H. Tajima, T. Tanaka, P. Turin, D. Glaser, J. Fermin, and R. P. Lin, "The focusing optics x-ray solar imager (FOXSI): instrument and first flight,", SPIE, 2013, vol. 8862, pp. 88620R–88620R–12, https://doi.org/10.1117/12.2024277

117     Y. E. Litvinenko and I. J. D. Craig, "Flare Energy Release by Flux Pile-up Magnetic Reconnection in a Turbulent Current Sheet," ApJ, vol. 544, no. 2, pp. 1101–1107, Dec. 2000, https://doi.org/10.1086/317262

118     W. Liu, V. Petrosian, B. R. Dennis, and G. D. Holman, "Conjugate Hard X-Ray
        Footpoints in the 2003 October 29 X10 Flare: Unshearing Motions, Correlations, and
        Asymmetries," ApJ, vol. 693, no. 1, pp. 847–867, Mar. 2009,
        https://doi.org/10.1088/0004-637X/693/1/847

119     W. Liu, Q. Chen, and V. Petrosian, "Plasmoid Ejections and Loop Contractions in an
        Eruptive M7.7 Solar Flare: Evidence of Particle Acceleration and Heating in
        Magnetic Reconnection Outflows," ApJ, vol. 767, no. 2, p. 168, Apr. 2013,
        https://doi.org/10.1088/0004-637X/767/2/168

120     D. W. Longcope and S. J. Bradshaw, "Slow Shocks and Conduction Fronts from
        Petschek Reconnection of Skewed Magnetic Fields: Two-fluid Effects," ApJ, vol.
        718, no. 2, pp. 1491–1508, Aug. 2010, https://doi.org/10.1088/0004-
        637X/718/2/1491

145     M. Oka, S. Krucker, H. S. Hudson, and P. Saint-Hilaire, "Electron Energy Partition in
        the Above-the-looptop Solar Hard X-Ray Sources," ApJ, vol. 799, no. 2, p. 129, Feb.
        2015, https://doi.org/10.1088/0004-637X/799/2/129

149     J. Qiu, H. Wang, C. Z. Cheng, and D. E. Gary, "Magnetic Reconnection and Mass
        Acceleration in Flare-Coronal Mass Ejection Events," ApJ, vol. 604, no. 2, pp. 900–
        905, Apr. 2004, https://doi.org/10.1086/382122

155     J. W. Reep, S. J. Bradshaw, and D. Alexander, "Optimal Electron Energies for
        Driving Chromospheric Evaporation in Solar Flares," ApJ, vol. 808, no. 2, p. 177,
        Aug. 2015, https://doi.org/10.1088/0004-637X/808/2/177

157     D. F. Ryan, W. H. Baumgartner, M. Wilson, A. BenMoussa, M. Campola, S. D.
        Christe, S. Gissot, L. Jones, J. Newport, M. Prydderch, S. Richards, P. Seller, A.
        Shih, and S. Thomas, "Tolerance of the High Energy X-ray Imaging Technology
        ASIC to potentially destructive radiation processes in Earth-orbit-equivalent
        environments," J. Inst., vol. 13, no. 2, pp. P02030–P02030, Feb. 2018,
        https://doi.org/10.1088/1748-0221/13/02/p02030

165     P. Seller, M. D. Wilson, M. C. Veale, A. Schneider, J. A. Gaskin, C. A. Wilson-
        Hodge, S. D. Christe, A. Shih, K. Gregory, A. R. Inglis, and M. Panessa, "CdTe focal
        plane detector for hard x-ray focusing optics," SPIE, 2015, vol. 9601, p. 960103,
        https://doi.org/10.1117/12.2191617

169     B. V. Somov and T. Kosugi, "Collisionless Reconnection and High-Energy Particle Acceleration in Solar Flares," ApJ, vol. 485, no. 2, pp. 859–868, Aug. 1997, https://doi.org/10.1086/304449

170     B. V. Somov and A. V. Oreshina, "Slow and fast magnetic reconnection. II. High-temperature turbulent-current sheet," Astronomy and Astrophysics, vol. 354, pp. 703–713, Feb. 2000, https://ui.adsabs.harvard.edu/link_gateway/2000A%26A...354..703S/ADS_PDF

171     L. Sui and G. D. Holman, "Evidence for the Formation of a Large-Scale Current Sheet in a Solar Flare," ApJ, vol. 596, no. 2, pp. L251–L254, Oct. 2003, https://doi.org/10.1086/379343

174     M. Temmer, A. Veronig, E. P. Kontar, S. Krucker, and B. Vršnak, "Combined STEREO/RHESSI Study of Coronal Mass Ejection Acceleration and Particle Acceleration in Solar Flares," ApJ, vol. 712, no. 2, pp. 1410–1420, Apr. 2010, https://doi.org/10.1088/0004-637X/712/2/1410

188     V. V. Zharkova, K. Arzner, A. O. Benz, P. Browning, C. Dauphin, A. G. Emslie, L. Fletcher, E. P. Kontar, G. Mann, M. Onofri, V. Petrosian, R. Turkmani, N. Vilmer, and L. Vlahos, "Recent Advances in Understanding Particle Acceleration Processes in Solar Flares," Space Science Reviews, vol. 159, no. 1, pp. 357–420, Sep. 2011, https://doi.org/10.1007/s11214-011-9803-y

203     Krucker et. al, "The Spectrometer/Telescope for Imaging X-rays (STIX)", A&A 642, A15 (2020), https://doi.org/10.1051/0004-6361/201937362

204     Allred, J. et al., "Modeling the Transport of Nonthermal Particles in Flares Using Fokker-Planck Kinetic Theory", ApJ, 902, page 14, 2020, https://doi.org/10.3847/1538-4357/abb239

| | Given name(s) | Family name(s) | ORCID iD | Email | Affiliation 1 | Affiliation 2 |
|---|---|---|---|---|---|---|
| Author 1<br>(no need to fill this column) | Steven | Christe | 0000-0001-6127-795X | steven.christe@nasa.gov | NASA Goddard Space Flight Center | |
| | Meriem | Alaoui | 0000-0003-2932-3623 | meriem.alaouiabdallaoui@nasa.gov | IREAP, University of Maryland | |
| | Joel | Allred | 0000-0003-4227-6809 | joel.c.allred@nasa.gov | NASA GSFC | |
| | Marina | Battaglia | 0000-0003-1438-9099 | marina.battaglia@fhnw.ch | University of Applied Sciences and Arts Northwest Switzerland | |
| | Wayne | Baumgartner | 0000-0002-5106-0463 | wayne.baumgartner@nasa.gov | NASA MSFC | |
| | J.C. | Buitrago-Casas | 0000-0002-8203-4794 | milo@ssl.berkeley.edu | University of California Berkeley | |
| | Amir | Caspi | 0000-0001-8702-8273 | amir@boulder.swri.edu | Southwest Research Institute | |
| | Bin | Chen | 0000-0002-0660-3350 | bin.chen@njit.edu | New Jersey Institute of Technology | |
| | Thomas | Chen | | | Columbia University | |
| | Brian | Dennis | 0000-0001-8585-2349 | brian.r.dennis@nasa.gov | NASA Goddard Space Flight Center | |
| | James | Drake | 0000-0002-9150-1841 | drake@umd.edu | University of Maryland, College Park | |
| | Lindsay | Glesener | 0000-0001-7092-2703 | glesener@umn.edu | University of Minnesota | |
| | Iain | Hannah | 0000-0003-1193-8603 | iain.hannah@glasgow.ac.uk | University of Glasgow | |
| | Laura A. | Hayes | 0000-0002-6835-2390 | laura.hayes@esa.int | European Space Agency (ESA) | |
| | Hugh | Hudson | 0000-0001-5685-1283 | hhudson@ssl.berkeley.edu | University of California Berkeley | University of Glasgow |
| | Andrew | Inglis | 0000-0003-0656-2437 | andrew.inglis@nasa.gov | NASA Goddard Space Flight Center | |
| | Jack | Ireland | 0000-0002-2019-8881 | jack.ireland-1@nasa.gov | NASA Goddard Space Flight Center | |
| | James | Klimchuk | 0000-0003-2255-0305 | James.A.Klimchuk@nasa.gov | NASA Goddard Space Flight Center | |
| | Adam | Kowalski | 0000-0001-7458-1176 | adam.f.kowalski@gmail.com | National Solar Observatory, Univ. Colorado | LASP |
| | Säm | Krucker | 0000-0002-2002-9180 | krucker@berkeley.edu | University of Applied Sciences and Arts Northwest Switzerland | |
| | Anna Maria | Massone | 0000-0003-4966-8864 | annamaria.massone@cnr.it | MIDA, Dipartimento di Matematica, Università di Genova, Genova, Italy | |
| | Sophie | Musset | 0000-0002-0945-8996 | sophie.musset@esa.int | ESA/ESTEC | |
| | Michele | Piana | 0000-0003-1700-991X | piana@dima.unige.it | MIDA, Dipartimento di Matematica, Università di Genova, Genova, Italy | INAF - Osservatorio Astrofisico di Torino, Torino, Italy |
| | Daniel | Ryan | 0000-0001-8661-3825 | daniel.ryan@fhnw.ch | University of Applied Sciences and Arts Northwest Switzerland | |
| | Albert | Shih | 0000-0001-6874-2594 | albert.y.shih@nasa.gov | NASA Goddard Space Flight Center | |
| | Astrid | Veronig | 0000-0003-2073-002X | astrid.veronig@uni-graz.at | University of Graz | |
| | Nicole | Vilmer | 0000-0002-6872-3630 | Nicole.Vilmer@obspm.fr | LESIA, Paris Observatory/PSL | |
| | Alexander | Warmuth | 0000-0003-1439-3610 | awarmuth@aip.de | Leibniz Institute for Astrophysics Potsdam | |
| | Stephen | White | 0000-0002-8574-8629 | stephen.white.24@us.af.mil | Air Force Research Laboratory | |